\documentclass{caosp309}

\usepackage{graphicx}

\usepackage{natbib}
\articleNo{x}
\pubyear{2025}
\volume{x}
\volnumber{x}
\firstpage{1}
\received{December 20, 2024}
\accepted{January 14, 2025}

\def\BibTeX{{\rm B\kern-.05em{\sc i\kern-.025em b}\kern-.08em
             T\kern-.1667em\lower.7ex\hbox{E}\kern-.125emX}}

\begin{document}

%
\htitle{Distribution of CVs in the Galaxy and Their Position in the HR Diagram }
\hauthor{R.Canbay}

\title{Distribution of Cataclysmic Variables in our Galaxy and Their Position in the HR Diagram in the Gaia Era}


%
%
\author{
        R.\,Canbay\orcid{0000-0003-2575-9892}
             }

%
\institute{
           \ Istanbul University, Institute of Graduate Studies in Science, Programme of Astronomy and Space Sciences, 34116, Istanbul, Turkey, \\
           \email{rmzycnby@gmail.com}}

\date{March 8, 2003}

\maketitle

\begin{abstract}
In this study, the distances of stellar systems classified as cataclysmic variables in the literature were determined by using the distance compiled from the \citet{BailerJones2021}. The spatial distributions of cataclysmic variables in the heliocentric Galactic coordinate system are obtained and their positions in the Hertzsprung-Russell (HR) diagram constructed from Gaia colors are discussed.

\end{abstract}


\section{Introduction}

\label{intr}
Cataclysmic variables (CVs) are binary star systems consisting of a white dwarf (WD) as the primary component and a low-mass main-sequence star that fills its Roche lobe. Matter is transferred from the secondary to the primary through a gas stream and an accretion disk. In magnetic CVs, such as those in the polar (P) subgroup, the strong magnetic field of the WD  completely prevents the formation of an accretion disk. Instead, material is channeled onto the WD through accretion columns and channels. In intermediate polars (IPs), the magnetic fields are weaker and insufficient to fully suppress the formation of an accretion disk, resulting in the formation of a disrupted, partial accretion disk influenced by the WD's magnetic field.

\citet{Townsley2002} showed that the position of CVs on the HR diagram \citep{Hertzsprung1911, Russell1914} is determined by their evolutionary parameters, including the temperature of the WD, the mass of the donor star, and the luminosity of the accretion disk. However, due to the limited sample size of CVs and the inherent challenges in determining their distances, an analysis of their absolute magnitude distribution has not been possible until recently. With the advent of Gaia, this limitation has been overcome. \citet{Abril2020} found that the CVs obtained from Gaia DR2 are distinctly clustered according to their orbital periods, while also being dispersed between the main sequence and WD regions of the HR diagram. The authors observed a relationship between orbital period, color, and absolute magnitude. This is, in fact, consistent with the expected behavior from the CV evolutionary model; as the orbital periods of evolved CVs shorten, the luminosity of the secondary star decreases, causing the WD and the accretion disk, which are already dominant, to become even more prominent. This applies to magnetic systems as well, though no disk structure is observed in those cases. \citet{Abrahams2022} used Gaia DR2 and EDR3 data to explore a new relationship between color ($G_{\rm BP}–G_{\rm RP}$), absolute magnitude ($M_{\rm G}$), and orbital period ($P_{\rm orb}$), as well as to investigate the period gap and angular momentum loss mechanisms in cataclysmic variables. 

\section{Data and Analysis}

\citet{Canbay2023} reported 4,149 systems with astrometric and photometric data in the Gaia DR3 data release and classified as CVs. The Gaia DR3 catalogue for CVs, including their equatorial coordinates $(\alpha, \delta)_{\rm J2000}$, their trigonometric parallaxes ($\varpi$) and Gaia magnitudes $G$, $G_{\rm BP}$, $G_{\rm RP}$, was obtained from \citet{Gaia2023}. Distances calculated using the \citet{BailerJones2021} Bayesian method based on Gaia DR3 trigonometric parallaxes were also included in the catalog. \citet{Canbay2023} compared the distances ($d=1000/\varpi$) of CVs calculated from Gaia's trigonometric parallaxes with the distances ($d_{\rm BJ}$) obtained by the \citet{BailerJones2021} method with respect to the $G$-apparent magnitude. The analysis showed that for systems with $G\leq 18.5$ mag, the scattering in the distance measurements is significantly smaller than for fainter systems. To minimize potential uncertainties in the distance estimates, the sample was restricted to $G\leq 18.5$ mag, resulting in the identification of 1,714 CV systems that satisfied this criterion. Starting with 1,714 CV systems, controls were performed based on the subtypes of the systems included in the analyses, and the system distances and apparent magnitude limits were updated. After excluding five systems as outliers, the analysis was carried out using a final sample of 1,709 systems.

Photometric data are affected by the interstellar medium. In this study, \citet{Schlafly2011} dust map was utilized to correct the photometric data for absorptions and reddening. By inputting Galactic coordinates ($l,b$), the absorption value $A_{\infty}(V)$ in the $V$-band valid up to the Galactic boundary was obtained from the dust map. Given that the distances of the CVs are known, the absorption value between the Sun and the stellar system was calculated using the relation by \citet{Bahcall1980}:
\begin{equation}
A_{\rm d}(V)=A_{\infty}(V)\Biggl[1-\exp\Biggl(\frac{-\mid d_{\rm BJ} \times\sin b\mid}{H}\Biggr)\Biggr],
\end{equation}
In this equation, $d_{\rm BJ}$ represents the distances of the CVs as determined by \citet{BailerJones2021}, $b$ denotes the Galactic latitude of the system, $H$ is the scale height of the dust, and $A_{\rm d}(V)$ is the reduced absorption value. For this study, the scale height of the dust was adopted as $H$=125pc \citep{Marshall2006}. The reduced color excess of the systems was determined using the relation $E_{\rm d}(B-V)=A_{\rm d}(V)/3.1$. The total absorptions in $G$, $G_{\rm BP}$, and $G_{\rm RP}$ bands were obtained by using relations given as follows: 
\begin{eqnarray}
A(G)=0.83627\times 3.1 E_{\rm d}(B-V), \nonumber \\ 
A(G_{\rm BP})=1.08337\times 3.1 E_{\rm d}(B-V), \\ 
A(G_{\rm RP})=0.63439\times 3.1 E_{\rm d}(B-V). \nonumber\\ 
\nonumber
\end{eqnarray}
The selective absorption coefficients in Equation (2) were taken from \citet{ Cardelli1989}. Using these coefficients, the total extinction values were determined, allowing for the calculation of de-reddened apparent magnitudes in the $G$, $BP$ and $RP$-bands, denoted as $G_{0}$, $G_{\rm BP}$ and $G_{\rm RP}$, respectively. The absolute magnitudes $M_{\rm G}$ of CVs were calculated using the distance modulus formula $G_{\rm 0}-M_{\rm G} = 5\times \log(d_{\rm BJ})-5$, where $d_{\rm BJ}$ is the distance obtained from \citet{BailerJones2021}. 

\section{Results}
The distribution of CVs, categorized into subgroups within the heliocentric rectangular Galactic coordinate system, is presented in Figure 1, with the corresponding results provided in Table 1.

\begin{figure}
\centerline{\includegraphics[width=0.5 \textwidth,clip=]{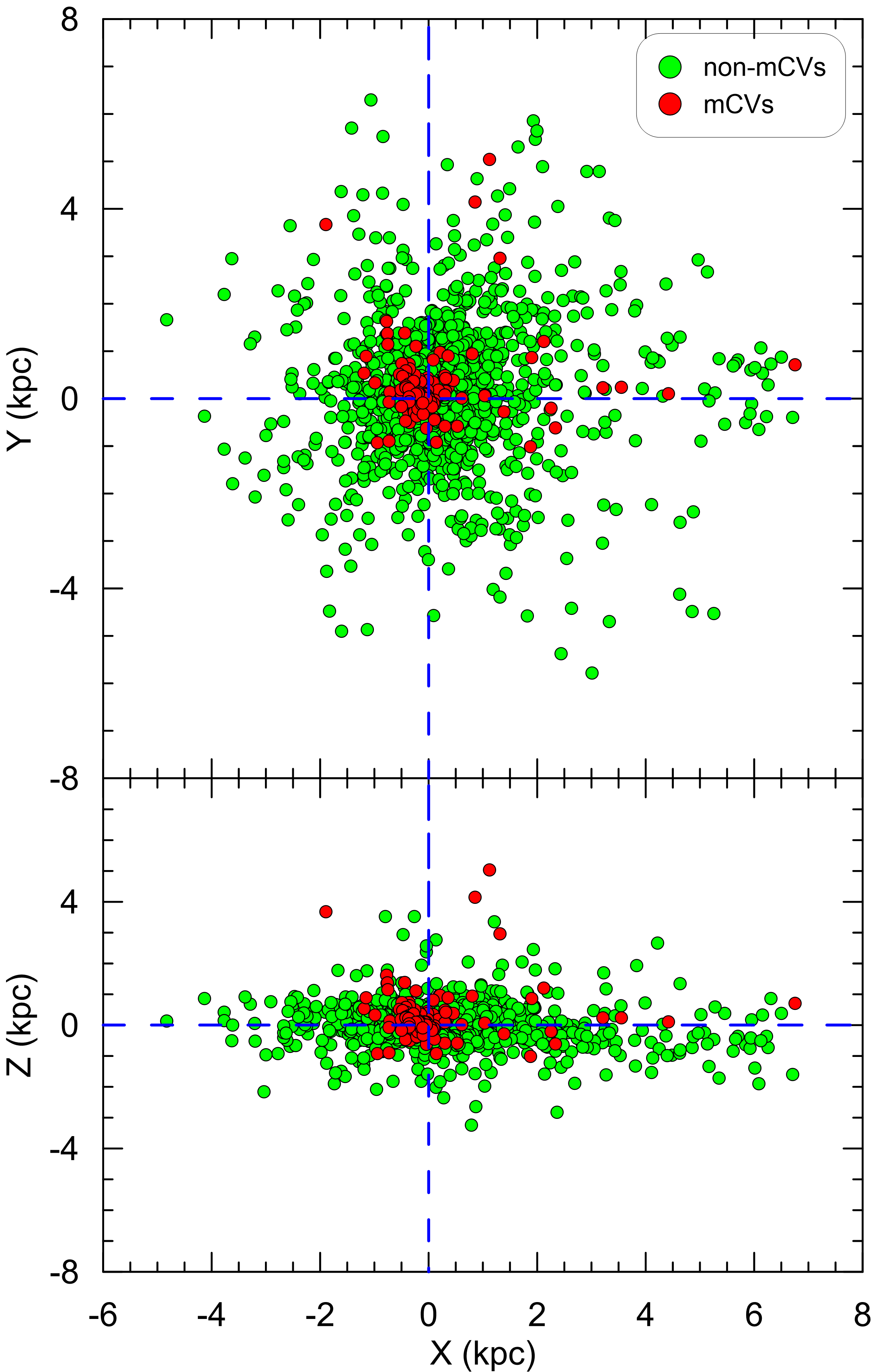}}
\caption{The spatial distribution of CVs with respect to the Sun. $X$, $Y$, and $Z$ are heliocentric 
 rectangular Galactic coordinates.}
\label{fsinus}
\end{figure}

\begin{figure}
\centerline{\includegraphics[width=0.5\textwidth,clip=]{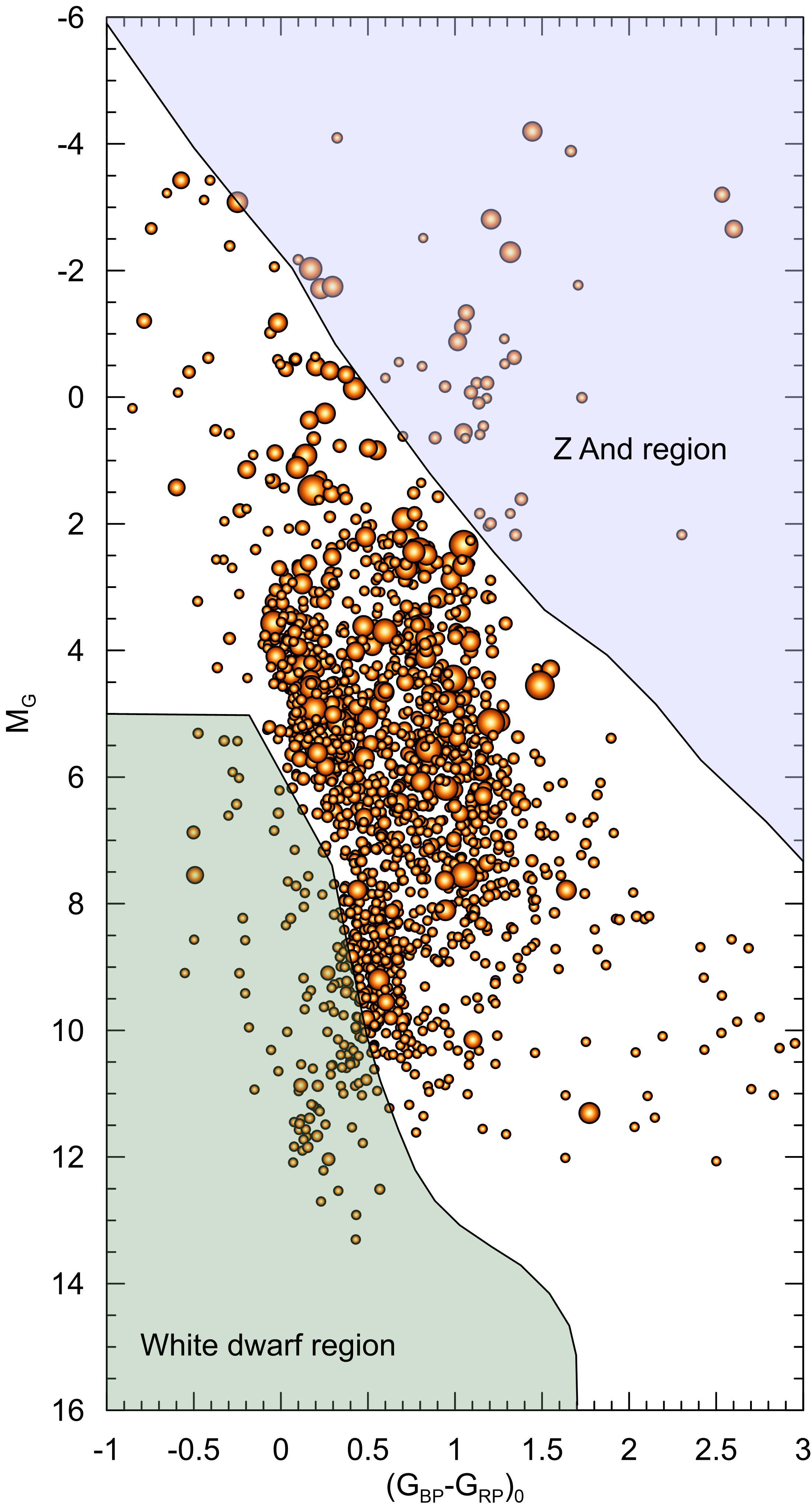}}
\caption{Distribution of known CVs in the HR diagram of Gaia DR3 database. The size of each point represents the relative parallax errors ($\sigma_{\rm\varpi}$/${\varpi}$) of all CVs.}
\label{fsinus}
\end{figure}

 The spatial distribution of the CVs used in this study reveals a concentration towards the Galactic center.  It displays trends similar to the Galactic coordinate distributions ($X$, $Y$, $Z$) reported by \citet{Canbay2023}. For the 1,587 CVs in the study, the median values of $d$ and ($X$, $Y$, $Z$) are 989 and (80, 93, -18) pc, respectively. These values are consistent with the median values found in the 1,709 samples analyzed in the study. The overall trends are consistent, with both studies showing a concentration of all CVs toward the Galactic center. Figure 2 shows their distribution based on absolute magnitude ($M_{\rm G}$) and color ($(G_{\rm BP}- G_{\rm RP})_{\rm 0}$). 

\begin{table}
\normalsize
\begin{center}
\caption{The median distances ($\tilde{d}$) and heliocentric rectangular Galactic 
coordinates ($\tilde{X}$, $\tilde{Y}$, $\tilde{Z}$) of CVs in the sample. Values are separately listed 
for All CVs, non-magnetic (non-mCVs) and magnetic (mCVs) systems. $N$ denotes the number of objects.}
\begin{tabular}{lccccc}
\hline 
\noalign{\vskip 0.1cm}
 Group      &   $N$  & $\tilde{d}$ & $\tilde{X}$  & $\tilde{Y}$  &  $\tilde{Z}$   \\ 
            &       & (pc) & (pc) & (pc) & (pc)   \\ 
 \hline
 All CVs        & 1,709  & ~~956  &  74  & 76  & -11 \\ 
 non-mCVs       & 1,577  & 1,003  &  92  & 93  & -13 \\ 
 mCVs           & ~~132  & ~~578  & -54  & -4  & ~19 \\ 
 \hline
\end{tabular}  
\end{center}
\end{table}

Considering the HR diagram obtained for CVs under the study, a focus on classifying them according to their components reveals a boundary separating WDs from main sequence stars, as identified by \citet{Fusillo2019}. Furthermore, the study by \citet{Gaia2019} analyzed the color-magnitude diagrams of variable stars, illustrating the distribution of subtypes, including Z Andromedae (Z And), U Geminorum (U Gem), and CVs. When applying this boundary to our dataset, it is observed that CVs are distributed across both the WD region, dominated by WDs and the Z And region, dominated by giant components. Objects situated between these two regions are inferred to exhibit a dominance of main sequence components (Figure 2). However, as shown in Figure 2, the majority of CVs are located in the main sequence (MS) region, suggesting that MS components are generally more prevalent. This finding aligns with the conventional definition of CVs, which consist of a WD and a low-mass MS star. Both components contribute significantly to the luminosity and color index of the system. Consequently, the observed colors and absolute magnitudes reflect the binary system as a whole, rather than the characteristics of each component individually. While the smaller, hotter WD dominates the luminosity at certain wavelengths, particularly in the ultraviolet, the MS star makes a substantial contribution to the system’s luminosity at optical wavelengths. Therefore, the classification and interpretation of CVs within the HR diagram should account for the combined properties of the binary system, rather than isolating the contributions of each component. 

 \citet{Abril2020} demonstrated that the distribution of CVs in the HR diagram forms distinct clusters corresponding to different stages of evolution. Their analysis revealed a trend where CVs with shorter orbital periods are fainter and exhibit bluer colors, a characteristic primarily influenced by the white dwarf’s luminosity and the declining contribution from the donor star.  These clusters align with standard evolutionary models of CVs. \citet{Abrahams2022} identified a relationship between the $(G_{\rm BP}- G_{\rm RP})$ color index, absolute magnitude, and orbital period. Their findings indicated that non-magnetic CVs here to a predictable pattern consistent with angular momentum loss mechanisms driving their evolution. Moreover, these studies underscore the significant influence of the accretion disk on the overall luminosity and color of CV systems. Our findings are consistent with these trends, as evidenced by the clustering of CVs in our HR diagram between the main sequence and white dwarf regions. When compared with the results of \citet{Abril2020} and \citet{Abrahams2022}, our data corroborate the observed color and magnitude transitions driven by the orbital period. The accretion disk significantly influences the luminosity and color index of CV systems. In non-magnetic CVs, the accretion disk’s luminosity can overshadow the contributions of the white dwarf and secondary star, especially during high accretion states. In contrast, in magnetic CVs such as polars, the absence of disk results in a lower overall luminosity, with the primary contributions coming from the white dwarf and accretion columns. A subset of CVs in our sample appears to exhibit luminosities characteristic of giant stars. This apparent discrepancy may arise from several factors; misclassification: The observational blending of CVs with background or nearby giant stars could artificially inflate their luminosity measurements. Spectroscopic confirmation is essential to disentangle such cases. Physical activity: In some CVs, the secondary star might exhibit enhanced activity, such as magnetic cycles or starspots, that temporarily boost the system’s brightness. Additionally, high accretion rates or transient outbursts could contribute to the observed luminosity. Further, spectroscopic and photometric investigations are essential to determine whether these systems are truly misclassified or if their elevated luminosities result from intrinsic activity. Refining these classifications will enhance our understanding of CV evolution and its distribution within the HR diagram.

\acknowledgements
This work has been supported in part by the Scientific 
and Technological Research Council of Turkey (T\"UB\.ITAK) 119F072. This work has been 
supported in part by Istanbul University: Project number NAP-33768. 

\bibliography{Canbay}

\end{document}